\documentclass{article}

\usepackage{arxiv}

\usepackage[utf8]{inputenc} % allow utf-8 input
\usepackage[T1]{fontenc}    % use 8-bit T1 fonts
\usepackage{hyperref}       % hyperlinks
\usepackage{url}            % simple URL typesetting
\usepackage{booktabs}       % professional-quality tables
\usepackage{amsfonts}       % blackboard math symbols
\usepackage{nicefrac}       % compact symbols for 1/2, etc.
\usepackage{microtype}      % microtypography
\usepackage{lipsum}
\usepackage{graphicx}
\graphicspath{ {./images/} }
\usepackage{bm}

\title{Reflection of Phase Anisotropy on Molecule}

\author{
 Yoshiaki Uchida \\
  Graduate School of Engineering Science\\
  Osaka University\\
  1-3 Machikaneyama, Toyonaka, Osaka 560-8531, Japan\\
  \texttt{y.uchida.es@osaka-u.ac.jp} \\
   \And
 Go Watanabe \\
  Department of Data Science, School of Frontier Engineering, \\
  Kitasato University\\
  Kanagawa 252-0373, Japan\\
  \texttt{go0325@kitasato-u.ac.jp} \\
}

\begin{document}
\maketitle
\begin{abstract}
This study explores how molecular shape changes influence the phase behavior of liquid crystals, particularly the nematic (N) phase of 5CB, through all-atom molecular dynamics (MD) simulations. The results demonstrate that molecular shape anisotropy increases in the N phase, with molecules adopting more elongated conformations as aggregation occurs. We find that the shape distribution is temperature- and aggregation-dependent, and the molecular shape relaxation time is longer in the N phase compared to isolated molecules. Additionally, the study proposes a revision to Onsager and Maier-Saupe’s theories, considering molecular shape distribution changes, which could improve the explanation of the N-I phase transition in liquid crystals. These findings contribute to a better understanding of the molecular behavior in liquid crystalline phases. 
\end{abstract}

\section{Introduction}
Now that liquid crystal displays have been widespread and commoditized, the physics of liquid crystalline (LC) phases can be well understood using the continuum theories like Landau-de Gennes theory \cite{1971degennes_ShortRangeOrderEffectsIsotropicPhaseNematicsCholesterics, 2002gennes_PhysicsLiquidCrystals, 2000emelyanenko_Moleculartheoryhelicalsenseinversionschiralnematicliquidcrystals}. The theory of LC phases and transitions can be used to explain various phenomena analogously, from the electronic and magnetic states of materials to biophysical properties. Some electronic and magnetic states have symmetries like LC phases due to their quantum effects \cite{2012togawa_ChiralMagneticSolitonLatticeChiralHelimagnet, 2021leonov_Surfaceanchoringcontrolparametershapingskyrmiontoronpropertiesthinlayerschiralnematicliquidcrystalsnoncentrosymmetricmagnets}. The theory of defects in LC phases \cite{2024fukuda_Directsimulationmachinelearningstructureidentificationunravelsoftmartensitictransformationtwinningdynamics, 2024zhao_Encapsulationtechniqueapplicationprogresspolydopaminebasedstimulusresponsivemicrocapsules, 2020yao_Electricallytunablepolarizationrandomlasingdyedopednematicliquidcrystals, 2025asilehan_Lightdrivendancingnematiccolloidsfractionalskyrmionsbimerons} has a common form with the theory that explains the process of biofilm formation \cite{2002gennes_PhysicsLiquidCrystals, 2023nijjer_Biofilmsselfshapinggrowingnematics}. The fluid dynamics theory of two-dimensional LC phases can also be used to explain cellular activities \cite{2022giomi_LongRangedOrderFlowAlignmentSheared$p$penalty1000hskip0ptaticLiquidCrystals, 2022giomi_Hydrodynamictheory$p$aticliquidcrystals, 2023armengol-collado_Epitheliaaremultiscaleactiveliquidcrystals}. The microscopic origins of their LC symmetries should be analogous to those of conventional LC materials. Understanding the molecular theory that is the basis of continuum theory should provide important insights into the fundamental causes of these diverse phenomena.

Molecular theories explain the microscopic origin of LC symmetries \cite{2001osipov_Helicaltwistingpowercirculardichroismnematicliquidcrystalsdopedchiralmolecules}. The molecular theory assumes the anisotropy of the intermolecular attraction and repulsion that determines the structure and properties of the nematic phase \cite{1931onsager_ReciprocalRelationsIrreversibleProcesses, 1959maier_einfachemolekularstatistischeTheorienematischenkristallinflussigenPhaseTeil, 1960maier_einfachemolekularstatistischeTheorienematischenkristallinflussigenPhaseTeilII}. Intuitively, it is more natural to think that anisotropic molecular shapes stabilize the LC phases by creating anisotropy in steric interactions. Onsager theory can explain lyotropic systems consisting of rigidly anisotropic particles. In contrast, the Maier-Saupe theory can explain thermotropic systems \cite{2001osipov_Helicaltwistingpowercirculardichroismnematicliquidcrystalsdopedchiralmolecules, 1977luckhurst_WhyMaierSaupetheorynematicliquidcrystalssuccessful}. To analyze the behavior of thermotropic nematic or cholesteric liquid crystals, it has become standard to use the Maier-Saupe form of molecular theory. However, given that new materials that exhibit new LC phases are still being discovered \cite{2017nishikawa_FluidLiquidCrystalMaterialHighlyPolarOrder, 2017mandle_nematicnematictransformationexhibitedrodlikeliquidcrystal}, it is difficult to say that the molecular science-based understanding that forms the basis of macroscopic continuum theory is sufficient. There are too many diverse macroscopic phenomena to be understood based on the picture of molecules as hard rod-like particles that interact with each other. Recently, it has been found that both the excluded volume effect and the attractive interaction are important in molecular dynamics (MD) simulations, and we are moving towards a more microscopic understanding \cite{2025ogita_Atomisticanalysisnematicphasetransition4cyano4nalkylbiphenylliquidcrystalsSamplingfirstorderphasetransitionfreeenergydecomposition}.

The LC symmetry reflects the symmetry of the constituent molecules \cite{2021mundoor_Thermallyreconfigurablemonoclinicnematiccolloidalfluids}. However, the molecules do not completely reflect their complex structures into LC phases. The symmetry of LC materials is generally higher than that of the constituent molecules. Intermolecular interactions are expected to play an important role in the hierarchical transformation of symmetry. When symmetry information is propagated in LC phases, shape similarity or complementarity between interacting molecules is important \cite{2022nemati_EffectsshapesolutesolventcompatibilityefficacychiralitytransferNanoshapesnematics}. Unlike crystals with three-dimensional periodicity, LC phases are difficult to observe experimentally at the atomic level. The molecular structure is generally discussed as being representative of the most stable conformation in a vacuum estimated by simulation or the conformation in a crystal structure measured experimentally. However, molecules should have a conformation ensemble at any point in time in LC phases, and the ensemble is continuously changing. While molecular softness has been theoretically discussed to explain the behavior of polymers, there are fewer examples of such theories for LC phases \cite{1979kimura_OrientationalPhaseTransitionSystemFlexibleMolecules, 2001toriyama_MechanicalStringModelLiquidCrystalEnsembleComprehensiveModelPolymorphism}. Recently, the importance of molecular softness, confirmed by machine learning studies \cite{2022uchida_ChemicalDataDrivenValidationPhysicalTheoriesLiquidCrystals}, has been brought to our attention.

Molecules in a vacuum have a conformational ensemble that follows the Boltzmann distribution. In contrast, a molecule in a crystal is left in one or a few conformations and has no other conformations. Using all-atom molecular dynamics (MD) simulations, we can obtain the structural time series data and investigate the conformational ensemble in a condensed phase.\cite{2023sadler_AutoregulationGPCRsignallingthirdintracellularloop, 2024sun_Subdomaindynamicsenablechemicalchainreactionsnonribosomalpeptidesynthetases} The construction of a theoretical framework that includes molecular fine structures requires the consideration of MD characteristics. The influence of the contact interactions could become more robust depending on the frequency of contacts, and the molecular conformational ensemble could change. This paper uses all-atom MD simulations to reveal how the molecular shape in liquid crystals is affected by contact interactions. We formulate the effect of the difference in molecular conformation ensemble between vacuum and LC phase. Besides, we demonstrate a framework for quantifying intermolecular contact interactions and predicting the behavior of the condensed phase by comparing MD simulations of single molecules and molecular clusters.

\begin{figure} % picture
  \centering
  \includegraphics[scale=0.5]{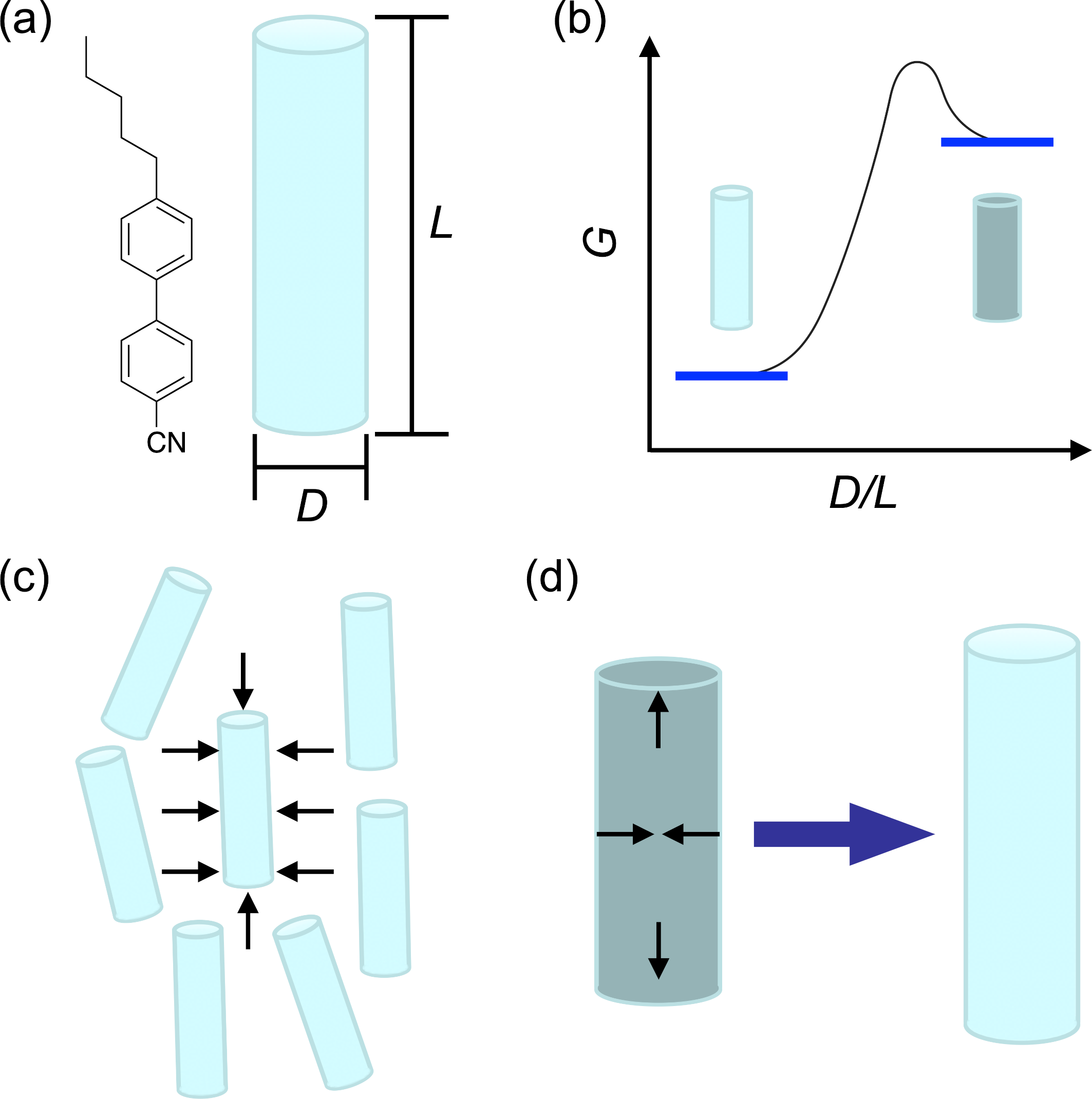}
  \caption{The proposed process by which molecules reflect the nematic symmetry. (a) Cylindrical approximation of a 5CB molecule showing a nematic phase. A molecule has a constant volume: $D^2 L=const.$ (b) The ensemble of conformations of 5CB on $D/L$ has bimodal distribution: one for elongated states and the other for short and thick states. (c) A molecule deforms anisotropically depending on the frequency of collisions with other molecules. (d) If the push from the side is strong, molecules with short and thick conformations change to elongated conformations.}
  \label{fig:fig1}
\end{figure}

\section{Methods}
\label{sec:headings}
\subsection{Molecular Dynamics Simulation}
All-atom MD simulation was carried out using the MD programs GROMACS 4.6.6 and 2016.3. The partial atomic charges of the simulated LC molecule were determined by the restrained electrostatic potential (RESP)\cite{1993bayly_wellbehavedelectrostaticpotentialbasedmethodusingchargerestraintsderivingatomicchargesRESPmodel} methodology at the UB3LYP/6-31G(d,p) level using the GAUSSIAN 09 package.\cite{2009m.j.frisch_Gaussian09RevisionD01} In order to calculate the intra- and intermolecular interactions, generalized Amber force field\cite{2004wang_Developmenttestinggeneralamberforcefield} parameters were used.

The initial structure of the MD simulation system was the cubic simulation box with a side of 10.99 nm containing 2009 molecules by replicating the small rectangular cell containing both enantiomers with random rotation. 4-cyano-4’-pentylbiphenyl was placed in the small rectangular unit cell. The relaxation runs at 250 K for 5 ns, at 350 K for 1 ns and 300 K for 5 ns were successively performed. During the relaxation runs, the Berendsen thermostat and barostat \cite{1984berendsen_Moleculardynamicscouplingexternalbath} were used to keep the temperature and pressure of the system with relaxation times of 0.2 and 2.0 ps, respectively. After the relaxation runs, the equilibration run at 300 K was done for 100 ns using Nos\'e-Hoover thermostat \cite{1984nose_moleculardynamicsmethodsimulationscanonicalensemble, 1985hoover_CanonicaldynamicsEquilibriumphasespacedistributions} and Parrinello-Rahman barostat \cite{1981parrinello_Polymorphictransitionssinglecrystalsnewmoleculardynamicsmethod} with relaxation times of 1.0 and 5.0 ps, respectively. The time step was set to 2 fs since all bonds connected to hydrogen atoms were constrained with the LINCS algorithm.\cite{1997hess_LINCSlinearconstraintsolvermolecularsimulations} The smooth particle-mesh Ewald (PME) method was employed to treat the long-range electrostatic interactions, and the real space cutoff and the grid spacing are 1.4 and 0.30 nm, respectively.

\section{Results and discussion}
\label{sec:headings}
\subsection{Equilibration of nematic phase of 5CB}
First, we simulate one of the most popular liquid crystalline compounds, 5CB, showing nematic (N) phases at 300 K and 350 K. The MD simulation of 2009 molecules of 5CB using Gromacs 2016.3. The 5CB molecules in the simulation at 300 K looked aligned in one direction. LC molecules tend to align along one direction after 600 ns of the equilibration run. For the quantitative analysis of the orientational order for LC molecules, we calculated the nematic order parameter. The nematic order parameter was estimated as the largest positive eigenvalue of the order parameter tensor $Q$, as expressed below:
\begin{equation}
Q_{\alpha \beta} = \frac{1}{N}\sum_{i=1}^N \frac{1}{2} \left( 3n_{i\alpha} n_{i\beta} - \delta_{\alpha \beta} \right)
\end{equation}
where $N$ is the total number of LC molecules, subscripts $\alpha$ and $\beta$ represent the coordinates $x$, $y$, and $z$, and $n_i$ is the normalized vector of the $i$th molecular long axis defined from the inertia tensor. We estimated $S$ as the largest positive eigenvalue of the order parameter tensor $Q$, as shown in Figure 2a. From the time series of $S$, we can assume that the system reached equilibrium after 600 ns. The average order parameter during the last 100 ns was $ 0.68  \pm 0.01$, suggesting that the system is in the N phase. The $S$ value is slightly higher than the experimentally obtained values. The mean square displacement (MSD) of the directions parallel and perpendicular to the director was analyzed for the 5CB molecules at 300 K. The MSDs were calculated to be $9.3 × 10^{-12} \mathrm{m}^2 \mathrm{s}^{-1}$ and $3.4 \times 10^{-12} \mathrm{m}^2 \mathrm{s}^{-1}$, which is comparable to the experimentally obtained values, as shown in Figure 2b, and the system has fluidity to the same extent as typical nematics \cite{2002dvinskikh_Anisotropicselfdiffusionthermotropicliquidcrystalsstudied$^1mathrmH$$^2mathrmH$pulsefieldgradientspinechoNMR}.

\begin{figure} % picture
  \centering
  \includegraphics[scale=0.6]{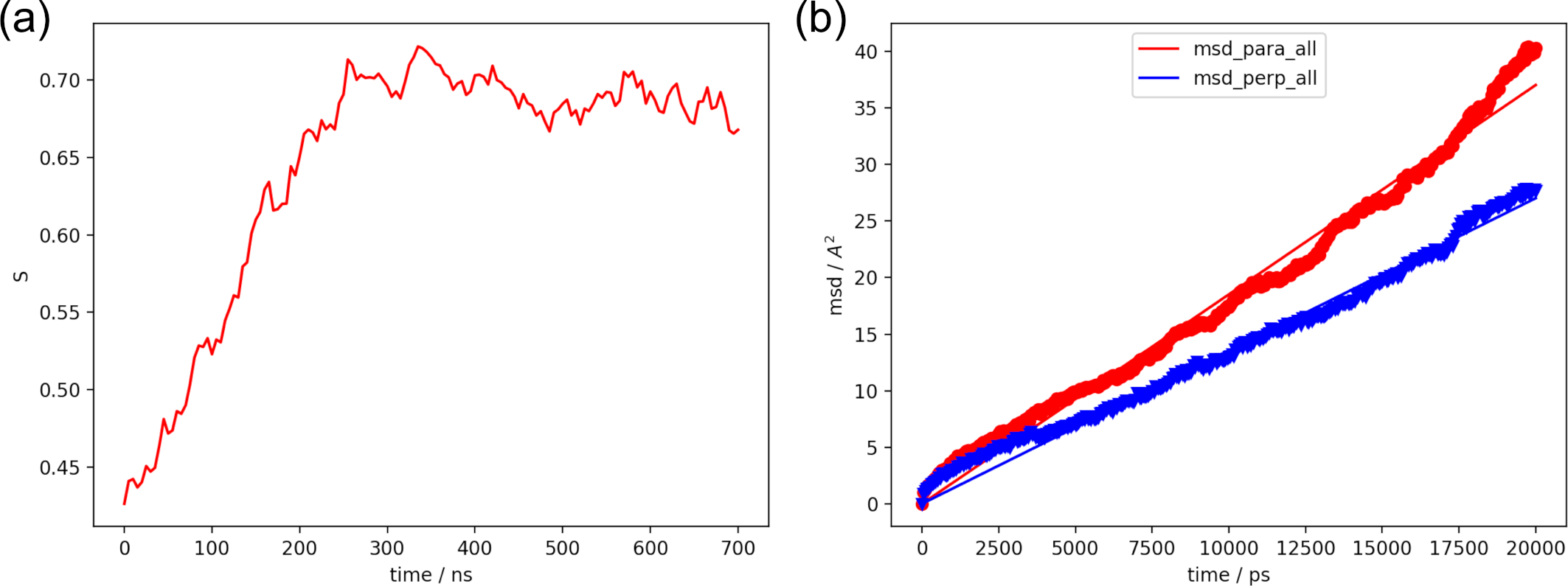}
  \caption{Equilibration process of 5CB at 300K and its diffusion anisotropy after the equilibration. (a) Time series of nematic scalar order parameter. (b) Mean square displacement of molecules in the N phase of 5CB. (c) Autocorrelation function of molecular uniaxiality in the N phase of 5CB.}
  \label{fig:fig2}
\end{figure}

The time constant of the autocorrelation function (ACF) of the molecular orientational direction at 300 K is defined as 
\begin{equation}
\rm{ACF} = \biggl\{\frac{1}{2}(3<\cos^2\theta_{lag}>-1)-\it{S}^{\rm{2}}\biggr\} / \left( 1- S^2 \right),
\end{equation}
where $\theta_{lag}$ is the lag of the orientation of each molecule. The ACF in the N phase for the MD simulation of 2009 5CB molecules at 300 K was fitted by the following equation.
\begin{equation}
\rm{ACF}=\exp\biggl\{{{-\biggr(\frac{\it{t}}{\rm{\tau}}\biggr)}^{\beta}}\biggr\}
\end{equation}
The time constant $\tau$ of ACF was estimated to be 12 ns, much longer than the molecular position (0.55 ns), as shown in Figure 3. It indicates that the orientation is rigid while it has high fluidity.

\begin{figure} % picture
  \centering
  \includegraphics[scale=0.6]{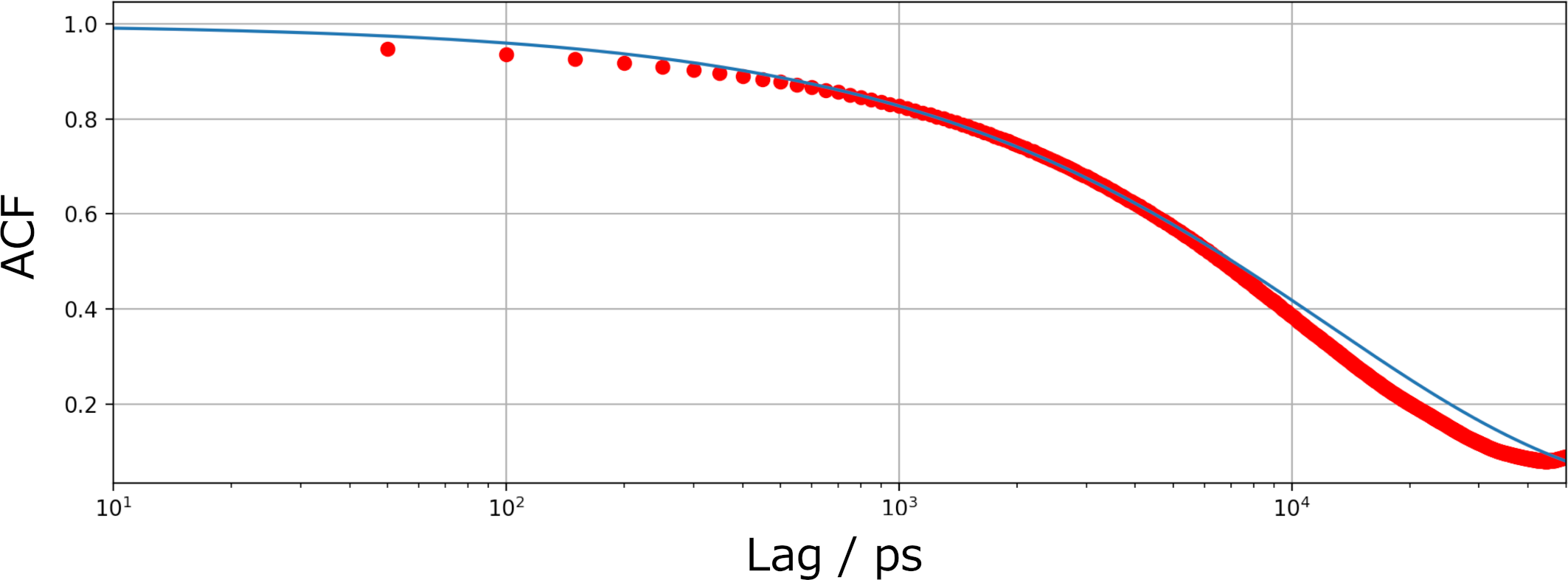}
  \caption{Autocorrelation function of molecular uniaxiality in the N phase of 5CB.}
  \label{fig:fig3}
\end{figure}

\subsection{Shape anisotropy of molecules}
We consider the molecular shape to be a cylinder corresponding to the single-axis symmetry of phase N, and discuss the anisotropy of the molecular shape. First, we calculated the inertia tensor of each molecule at each time from the results of MD simulations of the number of molecules at each temperature. The width ($D$) and length ($L$) of the molecules were approximated as cylinders, as shown in Figure 4a. First, in order to obtain the distribution that depends on the free energy of each conformation of the compound itself, 5CB, MD simulations were performed with a single molecule placed in a vacuum. In this case, a bimodal distribution was obtained. One mode was long and narrow, and the other mode was short and fat. This distribution is probably following the Boltzmann distribution.

What happens when molecules aggregate? The standard for the aggregation state should be the liquid state. Therefore, we performed MD simulations of 2009 molecules at 350K. As shown in Figure 4b, the peaks of the thick and short conformations have decreased, and the peaks of the elongated conformations have increased. For the liquid state, it is usually thought that the conformations are densely packed in the same way as in a vacuum state, but this is not the case. In other words, even if the molecules are simply in a condensed state, the ensemble of conformations may change due to collisions between molecules.

What happens in the liquid crystal state? The results of the MD simulation of 2009 molecules at 300 K are shown in Figure 4c. Looking at the probability distribution of $D/L$, we can see that the distribution is concentrated in a significantly elongated conformation compared to the liquid state. Furthermore, the half-width of the elongated peak has also narrowed. In the liquid crystal phase, the shape anisotropy of the molecules increases, and the shapes are also more uniform.

\begin{figure} % picture
  \centering
  \includegraphics[scale=0.8]{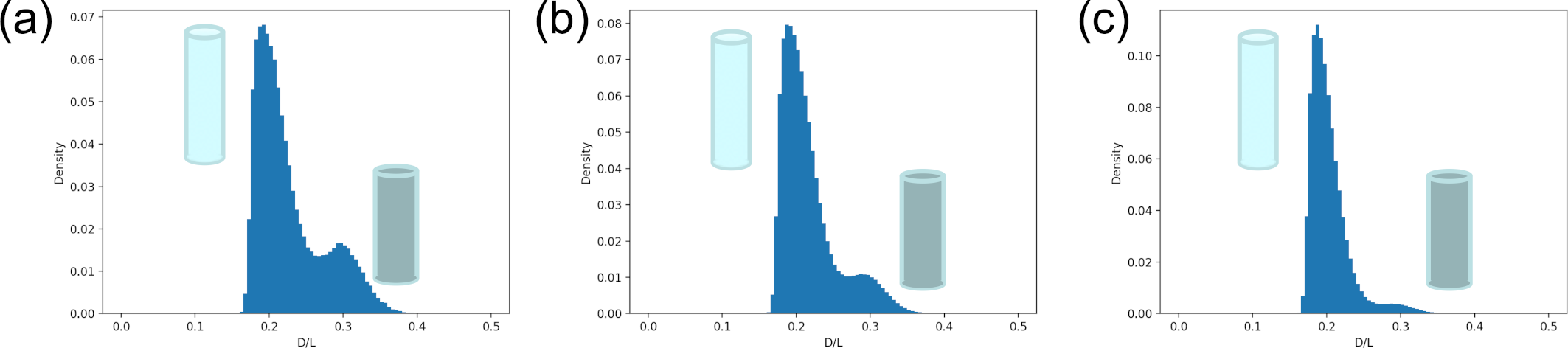}
  \caption{The conformation ensemble of molecules depends on the environment in which the molecules are placed. (a) The distribution of $D/L$ for the simulation of 1 molecule at 300 K. It is in a vacuum. (b) The distribution of $D/L$ for the simulation of 2009 molecules at 350 K. It is in a liquid phase and the number of molecules with higher $D/L$ decreases. (c) The distribution of $D/L$ for the simulation of 2009 molecules at 300 K. It is in a nematic phase and the molecules with higher $D/L$ almost disappear.}
  \label{fig:fig4}
\end{figure}

How does the aggregation of molecules affect molecular shape? We performed MD simulations at 300 K and 350 K for cases where 1, 8, 64, and 512 molecules were placed in a vacuum state and for cases where a continuous phase was created using a cell with 2009 molecules and periodic boundary conditions. As shown in Figure 5, it was found that the shape of the molecules depends on both temperature and the number of molecules that aggregate. First, it can be seen that the occupancy of the elongated shape is higher than that of the liquid phase at 350 K, which is the continuous phase for all numbers of molecules. This trend follows the Boltzmann distribution qualitatively. In addition, the results at 350 K, which can be explained only by the effect of aggregation due to the absence of the effect of orientation, are shown. As the aggregation of molecules above $10^2$ has the same distribution as the continuous phase, it can be seen that the effect of elongating molecules is saturated at around $10^2$.

Next, we will discuss the effect of the anisotropy of the nematic phase on the molecular shape. As we saw earlier, the reason why there are more elongated conformations than the temperature of the liquid is probably because this is more stable. Also, the rapid decrease in $D/L$ up to the aggregation of $10^2$ molecules must be due to the aggregation of molecules. On the other hand, even when the number of molecules that aggregate exceeds $10^2$ molecules, the decrease in $D/L$ does not stop. In the case of the 2009 molecule, this is thought to be the limit because it is a continuous phase, but it is expected that the number of molecules that aggregate to reach saturation is many orders of magnitude greater than in the liquid. This trend may correspond to the confinement effect, in which the phase transition temperature of the liquid crystal changes depending on the number of molecules in the aggregate state.

\begin{figure} % picture
  \centering
  \includegraphics[scale=0.4]{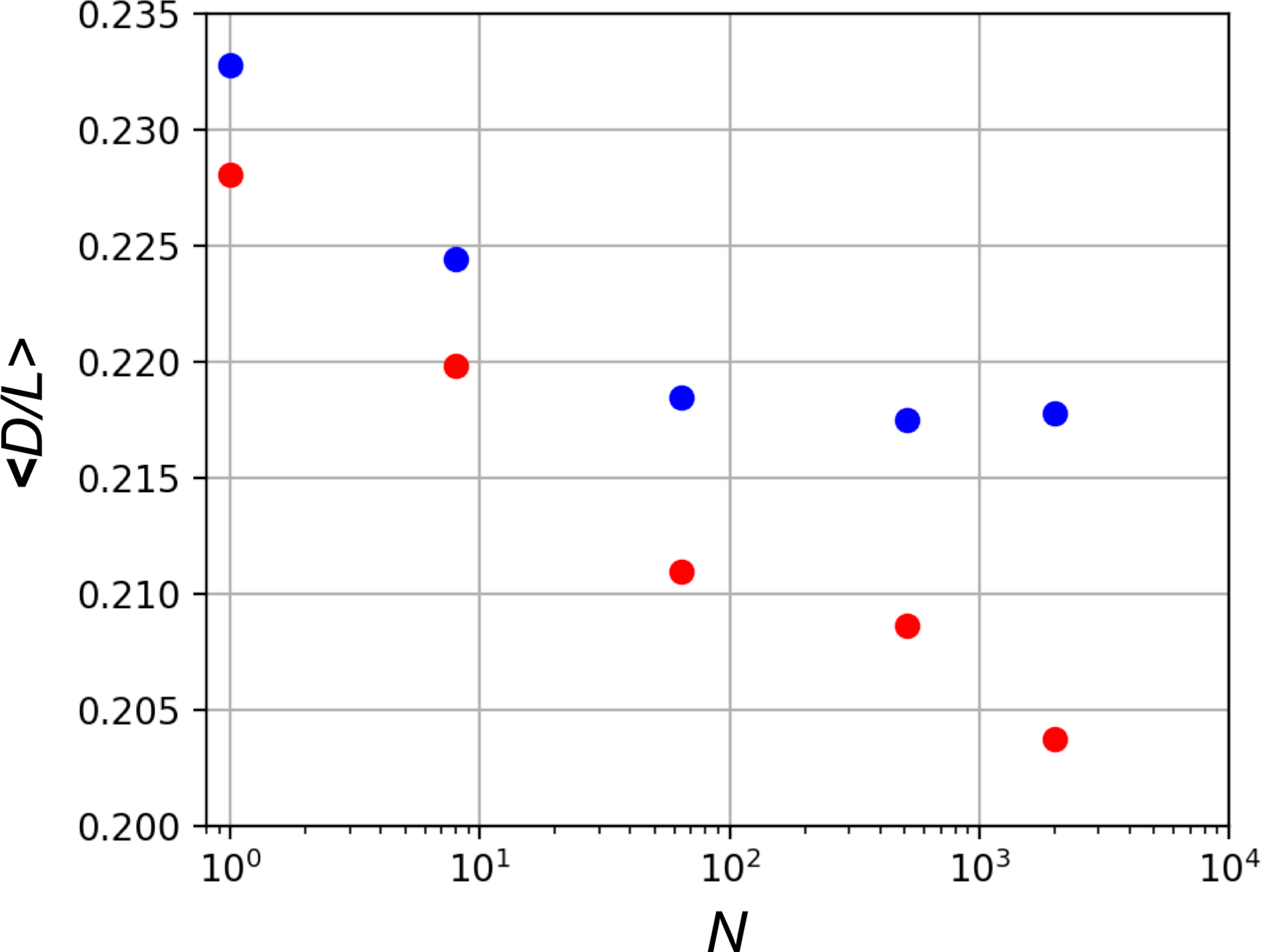}
  \caption{$\langle D/L\rangle$ - $N$ plot at 300 K (red dots) and 350 K (blue dots).}
  \label{fig:fig5}
\end{figure}

The effect of different aggregation states on molecular shape is discussed with different precision. We calculated the autocorrelation function of molecular shape. We fitted the autocorrelation function of D/L for the MD simulation of one 5CB molecule and 2009 5CB molecules at 300 K using equation 4, as shown in Figure 6. The estimated $\tau$ for the liquid phase is 547 ps, whereas the estimated $\tau$ for the liquid phase is 113 ps. The change in molecular shape is slowed down in the liquid crystal state by about 5 times. This may mean that the elongated molecular shape $D/L$ of the two original $D/L$ peaks is fixed around the $D/L$ of the elongated molecular shape.

\begin{figure} % picture
  \centering
  \includegraphics[scale=0.5]{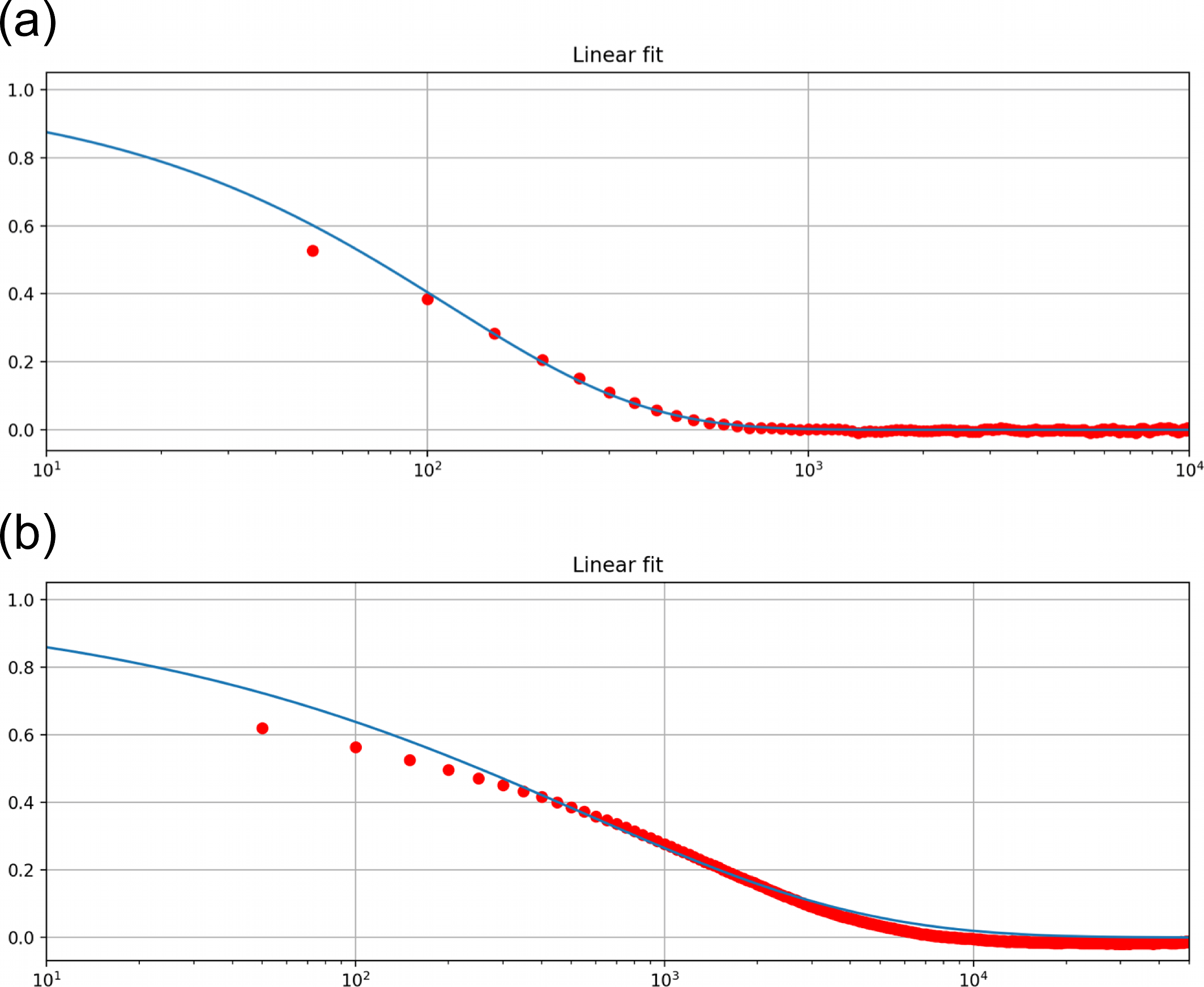}
  \caption{Autocorrelation function of $\langle D/L\rangle$ for the MD simulations (a) for one molecule and (b) for 2009 molecules at 300 K.}
  \label{fig:fig6}
\end{figure}

\subsection{The effect of molecular shape change on phase transition}

How would a change in molecular shape affect the phase transition? Molecular theory so far has not dealt explicitly with changes in molecular shape. The reason why Onsager's intuitively clear theory cannot explain the N-I phase transition well is that it assumes rigid molecules. If we consider that the phase transition changes the molecular shape, can we not explain it? In fact, it was recently confirmed that the excluded volume effect in Onsager's equation plays an important role, as does the gravitational interaction that appears in Maier-Saupe's equation \cite{2025ogita_Atomisticanalysisnematicphasetransition4cyano4nalkylbiphenylliquidcrystalsSamplingfirstorderphasetransitionfreeenergydecomposition}. For example, Onsager and Maier-Saupe theories for N phases involve nematic factor ($\Gamma_0$) in their free energy ($f_N$):
\begin{equation}
f_N=\frac{1}{2} \Gamma_0 \phi S^2- \int^1_0 \mathrm{exp} \left[\Gamma_0 S \left( \frac{3}{2}\xi^2 -\frac{1}{2} \right) \right]d\xi.
\end{equation}

For Onsager theory, the molecular length ($L$)-to-diameter ($D$) ratio ($L/D$) leads to nematic anisotropy, and the nematic factor is written as

\begin{equation}
\Gamma_0 = \frac{5}{4}\frac{L}{D}\phi
\end{equation}

where $\phi$ is the volume fraction, and $S$ is the nematic order parameter. Meanwhile, for Maier-Saupe theory, the nematic factor is written as

\begin{equation}
\Gamma_0 = \frac{U_{12}}{k_\mathrm{B} T}\frac{L}{D}\phi
\end{equation}

As shown in Table 1, the gravitational interaction $v_L = U_{12}/ k_\mathrm{B} T$ appearing in the Maier-Saupe equation is inversely proportional to temperature. On the other hand, the volume fraction$\phi$ appearing in Onsager's equation simply increases by a factor of 1.03 due to the effect of increasing density with decreasing temperature. Usually, thanks to this difference, Onsager's equation cannot quantitatively explain the N-I phase transition. In liquid crystals, however, the effect of molecular elongation appears in $L/D$. In fact, MD simulations show that it is 1.07 times larger at 300 K than at 350 K. $1.03 \times 1.07 = 1.10$, thus approaching the value of Maier-Saupe's $v_L$. Although more detailed analysis is needed in the future, if $L/D$ and $\phi$ change with temperature, Onsager's formula may be a better fit than Maier-Saupe's formula. Also, given that these depend on the number of molecules that aggregate, the gravitational interactions assumed by Maier-Saupe's equation are definitely affected, albeit indirectly.

\begin{table}
 \caption{Effects of molecular softness on N-I phase transition}
  \centering
  \begin{tabular}{cccc}
    \toprule
    $T$ & Attractive interaction ($v_L$) & Volume fraction ($\phi$) & Shape anisotropy ($L/D$) \\
    \midrule
    300 & 1.17 & 1.03 & 1.07\\
    350 & 1 & 1 & 1\\
    \bottomrule
  \end{tabular}
  \label{tab:table1}
\end{table}

\section{Conclusion}
\label{sec:others}
We found that 5CB is elongated in its N phase. Molecular shape distribution reflects the phase anisotropy. The molecular conformation ensemble depends on the environment in which the molecule is placed; it depends both on temperature and the number of molecules in a cluster. The molecular shape distribution may contribute to stabilizing the liquid crystalline phase. Furthermore, we found that the relaxation time of $\langle D/L\rangle$ in the N phase is larger than a single molecule. The contact between the molecules maintains the shape of the molecules. Besides, we proposed a new interpretation of Onsager and Maier Saupe's equations that the change in the molecular shape distribution affects the N-I phase transition, and the molecular shape depends on the molecular attraction interactions. This may be one way of connecting traditional molecular theories with organic chemistry.

\section{Acknowledgements}
\label{sec:others}
This work was supported in part by JSPS KAKENHI Grant Number JP23K23426. The computation was performed using Research Center for Computational Science, Okazaki, Japan (Project: 24-IMS-C050). 

\section{Disclosure statement}
\label{sec:others}
No potential conflict of interest was reported by the authors.

\bibliographystyle{unsrt} 
\bibliography{Myself}  

\end{document}